\documentclass[10pt,twocolumn,english,aps,prx,superscriptaddress,citeautoscript,nofootinbib]{revtex4-2}
\usepackage[T1]{fontenc}
\usepackage[utf8]{inputenc}
\usepackage{xcolor}
\usepackage{babel}
\usepackage{amsmath}
\usepackage{amssymb}
\usepackage{graphicx}
\usepackage{esint}
\usepackage[bookmarks=false,
 breaklinks=false,pdfborder={0 0 1},backref=section,colorlinks=true]
 {hyperref}
\hypersetup{
 linkcolor=blue,citecolor=blue,urlcolor=blue}

\makeatletter

\providecommand{\tabularnewline}{\\}

\makeatother

\begin{document}
\title{Operational modes of a Raman-coupled two-qubit quantum thermal machine}
\author{Alonso Alcalá}
\author{Charlie Oncebay}
\affiliation{Faculty of Sciences, National University of Engineering, 15032 Lima,
Peru}
\author{Onofre Rojas}
\author{Moises Rojas}
\affiliation{Department of Physics, Institute of Natural Science, Federal University
of Lavras, 37200-900 Lavras-MG, Brazil}
\begin{abstract}
We investigate a quantum thermal machine composed of two qubits coupled
through a Raman-induced exchange interaction and driven by inhomogeneous
transition frequencies. The system is analyzed within Carnot, Otto,
and Stirling thermodynamic cycles, including the Stirling cycle with
and without regeneration. We identify the conditions under which the
device operates as a heat engine, refrigerator, thermal accelerator,
or heater. Efficiency maps and operational-mode diagrams reveal well-defined
boundaries in parameter space, governed by the frequency ratio $r=\bar{\omega}/\omega$,
the coupling strength $g$, and the thermal gradient between reservoirs.
The Carnot cycle exhibits sharp transitions between engine and refrigerator
regimes, while the Otto cycle displays a richer structure with the
coexistence of all operational modes. The Stirling cycle shows enhanced
versatility and performance, particularly when assisted by a regenerator,
where near-ideal efficiencies are achieved. Overall, the Raman-type
interaction introduces a controllable left-right asymmetry that enables
nontrivial manipulation of thermodynamic behavior through frequency
tuning.
\end{abstract}
\maketitle

\section{Introduction}

Quantum thermodynamics investigates how familiar thermodynamic laws
emerge from quantum dynamics, and how coherence, entanglement and
measurement back-action modify energy conversion at the nanoscale
\cite{kos,cam,deff, binder,ali}. This framework raises fundamental questions about
the applicability of classical bounds, such as the Carnot limit, and
about the extent to which genuinely quantum resources can enhance
the performance of engines, refrigerators and related thermal devices
\cite{myers,moises}. Despite major progress, the field remains conceptually
open and continues to expand without a fully unified formulation \citep{bha}.

Quantum heat machines provide an operational setting in which these
issues can be probed. Cyclic protocols based on Hamiltonian modulation
and thermal contact have now been implemented in a wide range of platforms.
Single trapped ions have demonstrated fully controlled thermodynamic
cycles at the level of one particle \citep{ro}, while ultracold atomic
collisions allow heat exchange in quantized energy packets \citep{bou}.
Nuclear magnetic resonance experiments enable precise manipulation
of coupled spins with direct access to work and entropy production
\citep{pet}. In mesoscopic solid-state devices, quantum dots and
nanoelectronic conductors operate near thermodynamic limits \citep{jose,peko},
and Josephson-based elements support phase-coherent heat transport
and thermoelectric effects \citep{mar,gia}. Opto- and nanomechanical
setups use light-matter interactions or engineered noise to generate
refrigeration and heat flow \citep{zha,dech}. Circuit quantum electrodynamics
(circuit-QED) platforms, with superconducting qubits interacting through
microwave resonators, implement autonomous refrigeration as well as
measurement-based feedback protocols \citep{hofer,har}. These diverse
realizations collectively demonstrate that thermodynamic laws can
be investigated across many physical settings, each with different
degrees of control, noise and scalability.

Within this landscape, interacting spin systems occupy a prominent
role as versatile and experimentally accessible working media. Two-qubit
Heisenberg engines under static and anisotropic magnetic frequencies
have been investigated extensively, including effects of Dzyaloshinskii-Moriya
and Kaplan-Shekhtman-Entin-Wohlman-Aharony (KSEA) interactions \citep{he,alba,asa,kuz}.
Multi-spin extensions exploiting XXZ chains, Hubbard-type couplings
and controllable double quantum dots have broadened the range of architectures
where collective and correlated phenomena appear \citep{hee,moi,moi-1,oli}.
Spin-based refrigerators and Otto engines have also been studied near
critical points to assess the influence of many-body effects and entanglement
on performance \citep{cen,picci}. Stirling-type spin engines have
been shown to benefit from regeneration mechanisms that can approach
Carnot efficiency, and from anisotropic couplings that generate richer
operational behavior \citep{niu,ara,pili,ras,yin,wang,reza}. Additional proposals
include higher-dimensional lattices, dipolar-coupled nuclear spins
and binuclear molecular complexes serving as effective two-spin working
media \citep{pili-1,cak,cruz}. In these settings, entanglement has
been examined as a functional resource with consistent thermodynamic
interpretation \citep{cha,pili-2}.

Despite this extensive literature, most studies rely on symmetric
Heisenberg- or Ising-type exchange interactions. In contrast, comparatively
less attention has been devoted to left-right asymmetric Raman-type
couplings, such as $\sigma_{L}^{z}\sigma_{R}^{x}$, which arise naturally
in ferromagnetic and nuclear-spin environments and can be engineered
in platforms including cavity QED, trapped ions, circuit QED and optomechanics
\citep{kar}. Moreover, the majority of existing works address a single
cycle (typically Otto), without a unified comparison across different
cycles acting on the same microscopic working substance or a systematic
mapping of distinct operating modes heat engine, refrigerator, heater
and thermal accelerator, within the same parameter space.

The present work addresses these gaps. We consider a two-qubit system
with Raman-type coupling as a concrete and experimentally motivated
platform, and we systematically analyze three standard quantum cycles
Carnot, Otto and Stirling (with and without regeneration), under the
same Hamiltonian and control scheme. By constructing operational-mode
diagrams and performance maps as functions of frequency parameters,
we identify the conditions for optimized work extraction or cooling
and reveal thermodynamic signatures unique to asymmetric level mixing.

The remainder of the paper is organized as follows. In Sec. \ref{sec:2}
we introduce the Hamiltonian, thermalization processes and the general
structure of the quantum cycles. Sec. \ref{sec:3} describes the operational
modes and thermodynamic behavior of the Carnot, Otto and Stirling
engines. Sec. \ref{sec:4} presents the performance analysis, including
efficiency and coefficient-of-performance maps and the role of regeneration.
Finally, Sec. \ref{sec:5} summarizes our main conclusions and discusses
it implications.

\section{Microscopic model and thermodynamic framework}\label{sec:2}

\begin{figure}[h]
\includegraphics[scale=0.6]{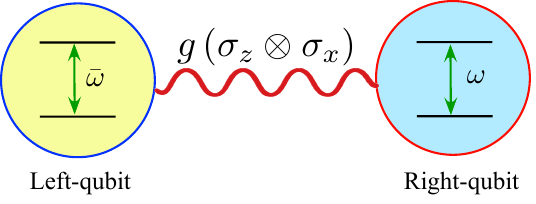} \caption{Schematic representation of the Raman-coupled two-qubit system, where
the right-qubit transition frequency is externally tuned while the
left qubit remains fixed.}\label{fig:dbl-qubit}
\end{figure}

We consider a system of two interacting qubits with transition frequencies
$\bar{\omega}$ and $\omega$, as illustrated in Fig. \ref{fig:dbl-qubit}.
This model, previously explored as a quantum optical two-atom thermal
diode \citep{kar}, is physically realized by spin-1/2 particles coupled
through an anisotropic Raman-type interaction. Throughout this work
we set $\hbar=1$. The system Hamiltonian is given by

\begin{equation}
H=\frac{\bar{\omega}}{2}(\sigma_{z}\otimes\mathbb{I})+\frac{\omega}{2}(\mathbb{I}\otimes\sigma_{z})+g\,(\sigma_{z}\otimes\sigma_{x}),
\end{equation}
 where $\sigma_{x}$ and $\sigma_{z}$ are Pauli matrices, $\bar{\omega}$
and $\omega$ denote the transition frequencies of the left and right
qubits, respectively, and $g$ is the coupling strength associated
with the Raman-induced exchange interaction.

In the computational basis $\{|00\rangle,|01\rangle,|10\rangle,|11\rangle\}$,
the Hamiltonian can be written in block-diagonal form as
\begin{equation}
H=\begin{pmatrix}H_{+} & 0\\
0 & H_{-}
\end{pmatrix},
\end{equation}
with two $2\times2$ blocks:
\begin{align}
H_{+} & =\begin{pmatrix}\frac{\bar{\omega}+\omega}{2} & g\\
g & \frac{\bar{\omega}-\omega}{2}
\end{pmatrix},\\
H_{-} & =\begin{pmatrix}-\frac{\bar{\omega}-\omega}{2} & -g\\
-g & -\frac{\bar{\omega}+\omega}{2}
\end{pmatrix}.
\end{align}
Diagonalizing these blocks yields four nondegenerate eigenvalues
\begin{equation}
\begin{aligned}E_{1} & =\frac{\bar{\omega}+\Omega}{2},\\
E_{2} & =\frac{\bar{\omega}-\Omega}{2},\\
E_{3} & =\frac{-\bar{\omega}+\Omega}{2},\\
E_{4} & =\frac{-\bar{\omega}-\Omega}{2},
\end{aligned}
\label{eq:E_i}
\end{equation}
where $\Omega=\sqrt{4g^{2}+\omega^{2}}$. 

The corresponding eigenvectors can be written as rotated basis states
with a mixing angle $\theta$, 
\begin{equation}
\begin{aligned}\left|\varphi_{1}\right\rangle  & =\cos\tfrac{\theta}{2}\left|00\right\rangle +\sin\tfrac{\theta}{2}\left|01\right\rangle ,\\
\left|\varphi_{2}\right\rangle  & =\sin\tfrac{\theta}{2}\left|00\right\rangle -\cos\tfrac{\theta}{2}\left|01\right\rangle ,\\
\left|\varphi_{3}\right\rangle  & =\sin\tfrac{\theta}{2}\left|10\right\rangle -\cos\tfrac{\theta}{2}\left|11\right\rangle ,\\
\left|\varphi_{4}\right\rangle  & =\cos\tfrac{\theta}{2}\left|10\right\rangle +\sin\tfrac{\theta}{2}\left|11\right\rangle ,
\end{aligned}
\end{equation}
where the angle satisfies $\tan(\theta)=\frac{2g}{\omega}$.

At thermal equilibrium with a bath at temperature $T$, the system
is described by the Gibbs state
\begin{equation}
\rho(T)=\frac{e^{-\beta H}}{Z}=\sum_{i=1}^{4}p_{i}\,|\varphi_{i}\rangle\langle\varphi_{i}|,
\end{equation}
where $\beta=1/(k_{B}T)$ is the inverse bath temperature and 
\begin{equation}
p_{i}=\frac{e^{-\beta E_{i}}}{Z},\qquad Z=\sum_{i=1}^{4}e^{-\beta E_{i}}\label{eq:p-Z}
\end{equation}
are the thermal populations and partition function, respectively.

Since the Hamiltonian is block-diagonal in the computational basis,
the thermal state preserves this structure: 
\begin{equation}
\rho(T)=\frac{1}{Z}\begin{pmatrix}\rho^{+} & 0\\
0 & \rho^{-}
\end{pmatrix},\label{eq:=00005Crho(T)}
\end{equation}
where $\rho^{+}$ acts in the $\{|00\rangle,|01\rangle\}$ subspace
and $\rho^{-}$ in the $\{|10\rangle,|11\rangle\}$ subspace. 
\begin{align}
\rho_{11}^{+} & =e^{-\beta\bar{\omega}/2}\left[\cosh\!\left(\tfrac{\beta\Omega}{2}\right)-\frac{\omega}{\Omega}\sinh\!\left(\tfrac{\beta\Omega}{2}\right)\right],\\
\rho_{22}^{+} & =e^{-\beta\bar{\omega}/2}\left[\cosh\!\left(\tfrac{\beta\Omega}{2}\right)+\frac{\omega}{\Omega}\sinh\!\left(\tfrac{\beta\Omega}{2}\right)\right],\\
\rho_{12}^{+} & =\rho_{21}^{+}=-\,\frac{2g}{\Omega}\,e^{-\beta\bar{\omega}/2}\sinh\!\left(\tfrac{\beta\Omega}{2}\right),\\[6pt]
\rho_{11}^{-} & =e^{\beta\bar{\omega}/2}\left[\cosh\!\left(\tfrac{\beta\Omega}{2}\right)-\frac{\omega}{\Omega}\sinh\!\left(\tfrac{\beta\Omega}{2}\right)\right],\\
\rho_{22}^{-} & =e^{\beta\bar{\omega}/2}\left[\cosh\!\left(\tfrac{\beta\Omega}{2}\right)+\frac{\omega}{\Omega}\sinh\!\left(\tfrac{\beta\Omega}{2}\right)\right],\\
\rho_{12}^{-} & =\rho_{21}^{-}=-\,\frac{2g}{\Omega}\,e^{\beta\bar{\omega}/2}\sinh\!\left(\tfrac{\beta\Omega}{2}\right),
\end{align}

These expressions reflect thermally induced coherence within each
subspace whenever $\theta\neq0$, evidencing the role of the Raman-type
interaction in mixing basis states even at equilibrium.

\subsection{Internal energy and entropy}

The thermodynamic properties of the system follow directly from the
Gibbs state in Eq. \eqref{eq:p-Z}. The internal energy is defined
as

\begin{equation}
U=\mathrm{Tr}\{\rho H\}=\sum_{i=1}^{4}p_{i}E_{i},
\end{equation}
where $E_{i}$ and $p_{i}$ are given in Eqs. \eqref{eq:E_i} and
\eqref{eq:=00005Crho(T)}, respectively. The von Neumann entropy of
the thermal state is
\begin{align}
\mathcal{S} & =-\mathrm{Tr}\{\rho\ln\rho\}=-k_{B}\sum_{i=1}^{4}p_{i}\ln p_{i}.
\end{align}
Since the spectrum depends on the control parameters $(\bar{\omega},\omega,g)$,
variations of these quantities change the internal energy according
to

\begin{equation}
dU=\sum_{i=1}^{4}E_{i}\,dp_{i}+\sum_{i=1}^{4}p_{i}\,dE_{i}.
\end{equation}
The first term accounts for population changes and corresponds to
energy exchange with the environment (heat). The second term stems
from modifications of the Hamiltonian eigenvalues and is associated
with the work performed on or by the system. This leads to the standard
decomposition used in quantum thermodynamics,

\begin{equation}
dU=\delta Q+\delta W,\label{eq:1st-law}
\end{equation}
where

\begin{equation}
\delta Q=\sum_{i}E_{i}\,dp_{i},\qquad\delta W=\sum_{i}p_{i}\,dE_{i}.\label{eq:dQ-dW}
\end{equation}
When the system remains in equilibrium with a bath at temperature
$T$ during a quasistatic transformation, the entropy change satisfies
the Clausius relation $d\mathcal{S}=\beta\,\delta Q$, ensuring full
thermodynamic consistency with the first and second laws.

\begin{figure}[h]

\includegraphics[scale=0.6]{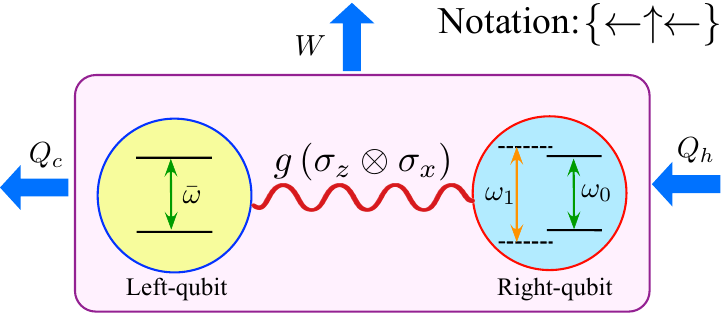} \caption{Visual depiction of the two-qubit working substance, where work is
produced by tuning the right-qubit transition frequency from $\omega_{0}$
to $\omega_{1}$.}\label{modos-maquina}

\end{figure}

\subsection{Thermodynamic strokes}\label{subsec:2B}

Quantum thermal machines operate through sequences of quasistatic transformations,
in which either the Hamiltonian parameters or the environmental conditions
are externally controlled. The working substance is a 
two-qubit system, schematically illustrated in Fig. \ref{modos-maquina}. In this setup,
useful work is generated by tuning the transition frequencies of the
right and left qubits, while heat arises
from population changes induced by thermal contact with the reservoirs.
Based on the decomposition introduced in Eq. \ref{eq:1st-law}, the relevant
thermodynamic strokes are defined as follows.

\paragraph{Isothermal process}

The system is weakly coupled to a thermal reservoir at fixed temperature
$T$. As the transition frequency is varied quasistatically, the state
remains Gibbsian, and both heat and work are exchanged according to
Eq. \eqref{eq:1st-law}. The entropy increases or decreases depending
on whether heat is absorbed from or released to the reservoir.

\paragraph{Constant-frequency (isochoric) process}

The transition frequencies are kept fixed, so that all energy eigenvalues
remain constant ($dE_{i}=0$). As a result, no work is performed,
$\delta W=0$. Heat exchange with the reservoir drives the populations
toward the corresponding thermal distribution, in accordance with
Eq. \eqref{eq:dQ-dW}.

\paragraph{Adiabatic process}

The system is isolated from the thermal reservoirs, and the evolution
is quasistatic, so that the instantaneous eigenstate populations remain
fixed ($dp_{i}=0$). Consequently, no heat is exchanged,
$\delta Q=0$, and Eq. \eqref{eq:dQ-dW} reduces to $dU=\sum_{i}p_{i}\,dE_{i}$.

The von Neumann entropy remains constant throughout the stroke. This
condition is intrinsically quantum: suppressing diabatic transitions
requires sufficiently slow modulation of the transition frequencies
($\bar{\omega},\omega$).

\subsection{Conditions modes}

When the system operates between a hot reservoir at temperature $T_{h}$
and a cold reservoir at temperature $T_{c}$, the net heat exchanged
with each bath and the total work performed over a complete cycle
satisfy $W=Q_{h}+Q_{c}$, together with the Clausius inequality
\begin{equation}
\oint\frac{\delta Q}{T}\le0,
\end{equation}
which restricts the physically admissible sign combinations of $(Q_{h},Q_{c},W)$.
As a consequence, only four distinct operation modes are allowed:
heat engine, refrigerator, thermal accelerator, and heater. Their
sign conventions and thermodynamic interpretation are summarized in
Table \ref{table:modesoperation}.

\begin{table}
\begin{tabular}{|c|c|c|c|c|c|}
\hline 
Mode & $Q_{c}$ & $W$ & $Q_{h}$ & Scheme & COP\tabularnewline
\hline 
\hline 
Engine & $-$ & $+$ & $+$ & $\{\leftarrow\uparrow\leftarrow\}$ & $\eta=\frac{W}{Q_{h}}$\tabularnewline
\hline 
Refrigerator & $+$ & $-$ & $-$ & $\{\rightarrow\downarrow\rightarrow\}$ & $\frac{Q_{h}}{W}$\tabularnewline
\hline 
Heater & $-$ & $-$ & $-$ & $\{\leftarrow\downarrow\rightarrow\}$ & $\frac{Q_{c}}{W}$\tabularnewline
\hline 
Accelerator & $-$ & $-$ & $+$ & $\{\leftarrow\downarrow\leftarrow\}$ & $\frac{Q_{c}}{W}$\tabularnewline
\hline 
\end{tabular}\caption{Sign conventions for the cold-bath heat $Q_{c}$, net work $W$,
and hot-bath heat $Q_{h}$ corresponding to the different operational
modes of the thermal machine.}\label{table:modesoperation}
\end{table}

The boundaries between these regimes depend on the transition frequencies
$(\bar{\omega},\omega)$, the coupling strength $g$, and the temperature
difference between the reservoirs. Mapping these boundaries in parameter
space constitutes the main objective of the analysis presented in
Sec. \ref{sec:4}.

\subsection{Performance measures}

In the heat engine regime, the performance is quantified by the thermal
efficiency
\begin{equation}
\eta=\frac{W}{Q_{h}},\qquad0\le\eta<1.
\end{equation}
 This quantity is bound by the second law and depends on the particular
sequence of thermodynamic strokes that form the cycle.

For the refrigerator, heater, and accelerator regimes, the most natural
metric is the coefficient of performance (COP), which may exceed unity.
To enable a direct comparison among all operational modes with a single
bounded parameter, we define the normalized performance parameter
\citep{moi} 
\begin{equation}
\kappa=\frac{\mathrm{COP}}{1+\mathrm{COP}},\qquad0<\kappa<1,
\end{equation}
which increases monotonically with efficiency. The limits of $\kappa$
provide a clear interpretation of performance: values near zero indicate
poor conversion efficiency, while values close to one correspond to
nearly ideal thermodynamic operation. This unified characterization
will be employed throughout our analysis of the Carnot, Otto, and
Stirling cycles in Sec. \ref{sec:3} and Sec. \ref{sec:4}.

\section{Quantum thermodynamic cycles }\label{sec:3}

The quantum Carnot, Otto, and Stirling cycles investigated in this study consist
of quasistatic stages, as described in Sec. \ref{subsec:2B}. In these cycles, useful
work is generated through the modulation of the qubit transition frequencies. Specifically, the cycle is driven by varying the frequency $\omega$ of the right qubit, while the frequency $\bar{\omega}$ of the left qubit is determined by the fixed 
ratio $r=\bar{\omega}/\omega$. This ratio quantifies the spectral asymmetry of the system, with $r=1$ corresponding 
to the symmetric configuration. Throughout this analysis, the frequency $\omega$ of the right qubit is varied between $\omega_{0}$ and 
$\omega_{1}$.  Heat  exchanged occurs via the redistribution of energy-level populations during the thermal
contact stages. The overall performance of each cycle depends on both the sequence of these stages 
and the possibility of internal heat recovery.

The following subsections summarize the structure of each cycle and
provide the basis for the operational maps presented in Sec. \ref{sec:4}.

\subsection{Carnot cycle}

The Carnot cycle consists of two isothermal and two adiabatic strokes.
Throughout the cycle, the right-qubit frequency is varied quasistatically
between $\omega_{0}$ and $\omega_{1}$.

All thermodynamic quantities are expressed in terms of the instantaneous
eigenvalues $E_{i}(\omega)$ and the corresponding thermal populations
$p_{i}(T,\omega)$. The cycle is initiated from the equilibrium state
at $(T_{c},\omega_{0})$, with populations $p_{i}^{(0)}=p_{i}(T_{c},\omega_{0})$.

\paragraph{Hot isothermal stroke $(\omega_{0}\to\omega_{1},\,T=T_{h})$:}

With the system in contact with the hot reservoir, the frequency is
swept from $\omega_{0}$ to $\omega_{1}$. The state remains Gibbsian
at temperature $T_{h}$ with final populations $p_{i}^{(h)}=p_{i}(T_{h},\omega_{1})$.
The heat absorbed from the hot bath follows from the entropy change
along the isotherm:
\begin{equation}
Q_{h}=T_{h}\,[S(\omega_{1},T_{h})-S(\omega_{0},T_{h})].
\end{equation}
Work is done through the variation of $E_{i}(\omega)$, but no explicit
formula is needed here since the Carnot efficiency is independent
of spectrum details.

\paragraph{Adiabatic expansion $(\omega_{1}\to\omega_{2})$:}

The system is isolated and the frequency changes from $\omega_{1}$
to $\omega_{2}$. The populations remain fixed at $p_{i}^{(h)}$ while
the eigenvalues change, so
\begin{equation}
Q_{2}=0,\qquad W_{2}=\sum_{i}p_{i}^{(h)}\,[E_{i}(\omega_{2})-E_{i}(\omega_{1})].
\end{equation}
This stroke preserves the entropy of the working medium.

\paragraph{Cold isothermal stroke $(\omega_{2}\to\omega_{3},\,T=T_{c})$:}

Coupled to the cold bath at $T_{c}$, the system returns to Gibbs
equilibrium as the frequency changes from $\omega_{2}$ to $\omega_{3}$.
The heat released is
\begin{equation}
Q_{c}=T_{c}\,[S(\omega_{3},T_{c})-S(\omega_{2},T_{c})],
\end{equation}
which is negative for an operating heat engine. Work is again associated
with the spectral variation.

\paragraph{Adiabatic compression $(\omega_{3}\to\omega_{0})$:}

In the final stroke the system is isolated and the frequency is returned
to its initial value $\omega_{0}$, with populations fixed at those
of the cold isotherm:
\begin{equation}
Q_{4}=0,\qquad W_{4}=\sum_{i}p_{i}^{(c)}\,[E_{i}(\omega_{0})-E_{i}(\omega_{3})].
\end{equation}
The state returns to $p_{i}^{(0)}$, closing the cycle.

\paragraph{Work and efficiency:}

The total work is $W=Q_{h}+Q_{c}.$ Under the quasistatic assumptions
used in this section, the efficiency attains its standard Carnot form,
$\eta_{\mathrm{Carnot}}=1-\frac{T_{c}}{T_{h}},$ independent of the
detailed spectral structure of the Raman-coupled two-qubit system.
The feasibility of engine or refrigeration operation for given $(\omega_{0},\omega_{1})$
depends instead on the signs of the entropy changes in the two isothermal
strokes, as analyzed in Sec. \ref{sec:4}.

\subsection{Quantum Otto cycle}

The Otto cycle consists of two adiabatic strokes and two constant-frequency
strokes. Analogously, the transition frequency of the right qubit is varied quasistatically between
$\omega_{0}$ and $\omega_{1}$.

The cycle starts from thermal equilibrium at $(T_{c},\omega_{0})$,
with populations $p_{i}^{(0)}=p_{i}(T_{c},\omega_{0})$.

\paragraph{First adiabatic stroke $(\omega_{0}\to\omega_{1})$:}

The system is isolated and the right-qubit frequency is changed from
$\omega_{0}$ to $\omega_{1}$. The populations remain fixed at $p_{i}^{(0)}$,
so this stroke exchanges only work 
\begin{equation}
Q_{1}=0,\qquad W_{1}=\sum_{i}p_{i}^{(0)}\,[E_{i}(\omega_{1})-E_{i}(\omega_{0})].
\end{equation}

\paragraph{Hot constant-frequency stroke at $\omega_{1}$:}

With the frequency fixed at $\omega_{1}$, the system is brought into
contact with the hot reservoir at temperature $T_{h}$ and relaxes
to the Gibbs state with populations $p_{i}^{(h)}=p_{i}(T_{h},\omega_{1})$.
Since the spectrum does not change, this stroke exchanges only heat,
\begin{equation}
W_{2}=0,\qquad Q_{h}=\sum_{i}E_{i}(\omega_{1})\,[p_{i}^{(h)}-p_{i}^{(0)}].
\end{equation}

\paragraph{Second adiabatic stroke $(\omega_{1}\to\omega_{0})$:}

The system is again isolated and the frequency is returned from $\omega_{1}$
to $\omega_{0}$. The populations remain equal to $p_{i}^{(h)}$,
so 
\begin{equation}
Q_{3}=0,\qquad W_{3}=\sum_{i}p_{i}^{(h)}\,[E_{i}(\omega_{0})-E_{i}(\omega_{1})].
\end{equation}

\paragraph{Cold constant-frequency stroke at $\omega_{0}$:}

Finally, at fixed frequency $\omega_{0}$ the system is coupled back
to the cold reservoir at $T_{c}$ and relaxes from $p_{i}^{(h)}$
to the initial populations $p_{i}^{(0)}$. This stroke exchanges only
heat, 
\begin{equation}
W_{4}=0,\qquad Q_{c}=\sum_{i}E_{i}(\omega_{0})\,[p_{i}^{(0)}-p_{i}^{(h)}].
\end{equation}
The net work and the first law follow directly, 
\begin{equation}
W=W_{1}+W_{3}=Q_{h}+Q_{c},
\end{equation}
and the efficiency in engine mode is
\begin{equation}
\eta_{\mathrm{Otto}}=1-\frac{|Q_{c}|}{Q_{h}},
\end{equation}
with all dependence on the control parameters entering through the
eigenvalues $E_{i}(\omega)$ and the thermal populations $p_{i}(T,\omega)$.
This structure underlies the operational maps and performance diagrams
in the $(\omega_{0},\omega_{1})$ plane discussed in Sec. \ref{sec:4}.

\subsection{Quantum Stirling cycle}

The Stirling protocol consists of two isothermal processes and two constant-frequency
processes. As in the previous cycles, the frequency of the right qubit is modulated between the values $\omega_{0}$
and $\omega_{1}$.

We first consider the conventional Stirling cycle without regeneration,
followed by its regenerated version.

\subsubsection{Stirling cycle without regeneration}

The cycle begins at $(T_{h},\omega_{0})$ with equilibrium populations
$p_{i}^{(h,0)}=p_{i}(T_{h},\omega_{0})$.

\paragraph{Hot isothermal stroke $(\omega_{0}\to\omega_{1},T=T_{h})$:}\label{par:Stirling-1.1}

The system is in contact with the hot bath while the frequency changes
from $\omega_{0}$ to $\omega_{1}$. The state remains Gibbsian with
populations $p_{i}^{(h,1)}=p_{i}(T_{h},\omega_{1})$. The heat absorbed
is 
\begin{equation}
Q_{h}=T_{h}\,[S(\omega_{1},T_{h})-S(\omega_{0},T_{h})].
\end{equation}
Work is generated through the variation of the eigenvalues.

\paragraph{First constant-frequency stroke at $\omega_{1}$:}

The coupling to the thermal reservoir is then switched to the lower temperature $T_{c}$, while the frequency is kept fixed at $\omega_{1}$. Since no spectral variation occurs during this stage, the process takes place at constant frequency. No work is performed, and heat is released during this process, denoted by $Q_{2}$
\begin{equation}
Q_{2}=\sum_{i}E_{i}(\omega_{1})\,[p_{i}^{(c,1)}-p_{i}^{(h,1)}],
\end{equation}
where $p_{i}^{(c,1)}=p_{i}(T_{c},\omega_{1})$ and $p_{i}^{(h,1)}=p_{i}(T_{h},\omega_{1})$. 

\paragraph{Cold isothermal stroke $(\omega_{1}\to\omega_{0},T=T_{c})$:}\label{par:Stirling-1.3}

The frequency is slowly returned from $\omega_{1}$ to $\omega_{0}$
while the system thermalizes at $T_{c}$. The heat released is
\begin{equation}
Q_{c}=T_{c}\,[S(\omega_{0},T_{c})-S(\omega_{1},T_{c})]<0.
\end{equation}

\paragraph{Second constant-frequency stroke at $\omega_{0}$:}

With the frequency fixed at $\omega_{0}$, the system completes its
relaxation back to $p_{i}^{(h,0)}$. As in the first constant-frequency
stroke, no spectral work is performed. The heat exchanged during this process is given by 
\begin{equation}
Q_{4}=\sum_{i}E_{i}(\omega_{0})\,[p_{i}^{(h,0)}-p_{i}^{(c,0)}],
\end{equation}
where $p_{i}^{(c,0)}=p_{i}(T_{c},\omega_{0})$. This final thermodynamic stroke simply
restores the system to its initial state.

\paragraph{Work and efficiency:}

The total work per cycle is
\begin{equation}
W=Q_{in}+Q_{out},
\end{equation}
where $Q_{in}=Q_{h}+Q_{4}$ and $Q_{out}=Q_{2}+Q_{c}$
and the Stirling efficiency follows from
\begin{equation}
\eta_{\mathrm{St}}=\frac{W}{Q_{h}+Q_{4}}.
\end{equation}
Because one of the two non-isothermal strokes is constant-frequency
with no spectral contribution, the achievable efficiency depends sensitively
on the entropy changes along the two isothermal legs, this dependence
is reflected in the operational maps of Sec. IV.

\subsubsection{Stirling cycle with regeneration}

In the regenerated Stirling cycle, the sequence of thermodynamic strokes is identical to that of the standard Stirling machine.
However, during the two constant-frequency strokes, heat is no longer exchanged with external reservoirs. Instead, the heat released in the first constant-frequency process is stored internally by the regenerator and subsequently reused in the second one, thereby suppressing external dissipation and enhancing the reversibility of the cycle.
The heats associated with the two isochoric strokes at $\omega_{1}$ and $\omega_{0}$ are denoted by $Q_{2}$ and $Q_{4}$, respectively, and their net contribution is

\begin{equation}
\Delta Q=Q_{2}+Q_{4}.\label{eq:del}
\end{equation}

As a consequence, the presence of the regenerator modifies the amount of heat effectively absorbed from the hot reservoir. 
The total heat input per cycle is then given by

\begin{equation}
Q_{in}^{reg}=Q_{h}+\delta \Delta Q.\label{eq:del-1}
\end{equation}
where $Q_{h}$ is the heat absorbed during the hot isothermal stroke and $\delta=1$ if $\Delta Q>0$, and $\delta=0$ if $\Delta<0$.
In the following, we restrict our analysis to the case $\Delta Q>0$, for which the regenerator effectively supplies heat to the cycle. When $\Delta Q<0$, the regenerator would instead require an additional heat input from the hot reservoir, offering no thermodynamic advantage and reducing the overall efficiency of the engine.

%\begin{equation}
%Q_{\mathrm{rel}}^{(\omega_{1})}=\sum_{i}E_{i}(\omega_{1})\,[p_{i}^{(c,\omega_{1})}-p_{i}^{(h)}].\label{eq:Qre1}
%\end{equation}

\paragraph{Work and efficiency:}

The total  output remains
\begin{equation}
W=Q_{in}+Q_{out}.
\end{equation}
with $Q_{out}=Q_{2}+Q_{c}$, and the efficiency of the Stirling engine with regenerator is corresponding defined as
\begin{equation}
\eta_{\mathrm{St}}^{\mathrm{reg}}=\frac{W}{Q_{in}^{reg}}.
\end{equation}
Due to the internal recovery of heat provided by the regenerator, the efficiency approaches its ideal limit. This mechanism accounts for the 
high-efficiency regions observed in the operational maps of the regenerated Stirling cycle presented in Sec. IV.

\section{Results and Discussions}\label{sec:4}

In this section, we investigate the operational modes and performance of
a Raman-coupled two-qubit system. The thermodynamic cycles are driven
by modulating the frequency $\omega$ of the right qubit, which is varied between $\omega_{0}$
and $\omega_{1}$. The parameter $r$ characterizes the spectral asymmetry of the system, with $r=1$ corresponding
to the symmetric configuration.

\subsection{Operational-mode diagrams}

For fixed temperatures and coupling, the Raman interaction and the
inhomogeneous frequencies transition $(\omega_{0},\omega_{1})$ control
the level structure and, consequently, the regions where the cycle
behaves as an engine, refrigerator, heater, or accelerator. Figures
3 to 6 display these modes for the Carnot, Otto, and Stirling cycles,
with and without regenerator, and illustrate how the choice of protocol
reshapes the operational boundaries in the $(\omega_{0},\omega_{1})$
plane.

\subsubsection{Quantum Carnot cycle}

\begin{figure}
\includegraphics[scale=0.43]{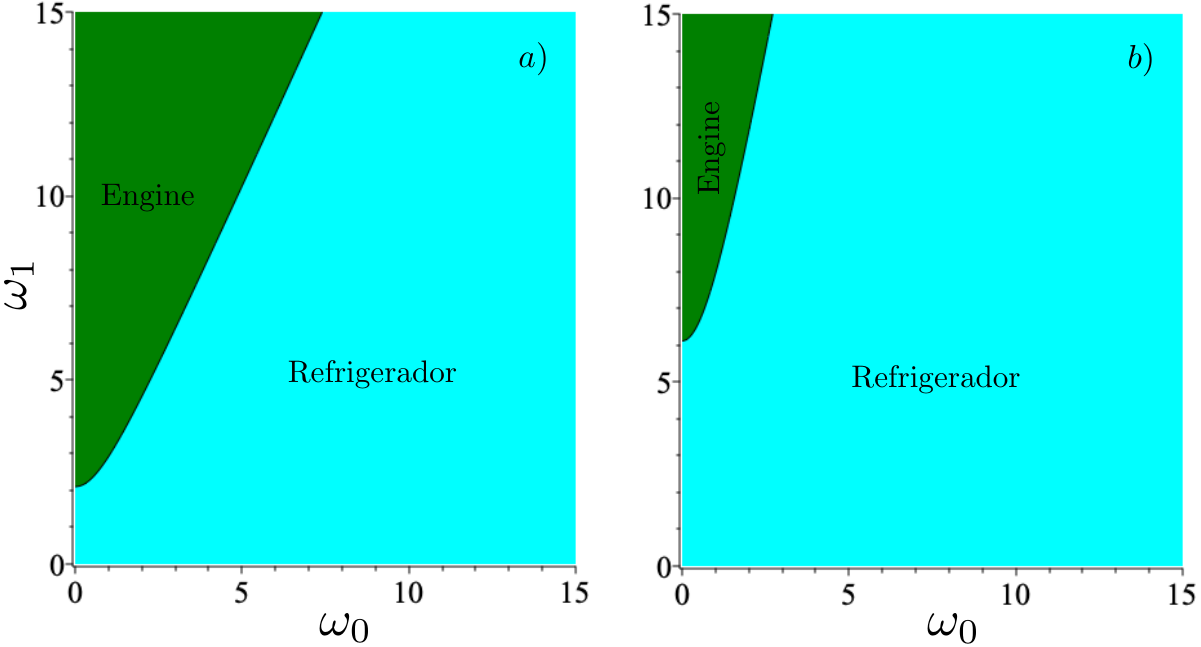} \caption{Schematic representation of Quantum Carnot machine cycle in the $\omega_{0}-\omega_{1}$
plane. a) For $T_{c}=1$, and $T_{h}=2$. b) For $T_{c}=1$, and $T_{h}=5$.
The parameters are set to $r=1$, and $g=1$.}
\label{fig:carnot} 
\end{figure}

Figure\textcolor{purple}{{} \ref{fig:carnot}} shows the operational-mode
diagrams of the quantum Carnot cycle in the $(\omega_{0},\omega_{1})$
plane. Here, $\omega_{0}$ and $\omega_{1}$ denote the two values
of the right-qubit frequency used during the adiabatic strokes, while
the left-qubit frequency is kept fixed. The colored regions indicate
whether the machine operates as a heat engine or as a refrigerator. 

In panel \ref{fig:carnot}(a) the reservoir temperatures are $T_{c}=1$
and $T_{h}=2$. Two distinct operating domains are clearly visible.
The green region corresponds to engine operation, where positive work
is extracted from the temperature difference between the baths. The
cyan region represents the refrigerator regime, in which external
work is required to transfer heat from the cold reservoir to the hot
one. The curve separating these regions corresponds to the condition
of zero net work over a complete cycle.

Panel \ref{fig:carnot}(b) displays the same cycle for a stronger
thermal gradient, with $T_{c}=1$ and $T_{h}=5$. In this case the
heat-engine region becomes narrower and shifts toward lower values
of $\omega_{0}$, while the refrigeration domain expands and occupies
most of the $(\omega_{0},\omega_{1})$ plane. This shows that, although
the Carnot efficiency increases with the ratio $T_{h}/T_{c}$, the
set of frequency parameters that support reversible work extraction
becomes more restricted when the temperature difference is large.

In summary, the Carnot cycle attains the maximal theoretical efficiency
wherever it operates as a heat engine, but the strict reversibility
conditions confine this regime to a relatively small region.

\subsubsection{Quantum Otto cycle}

\begin{figure}
\includegraphics[scale=0.41]{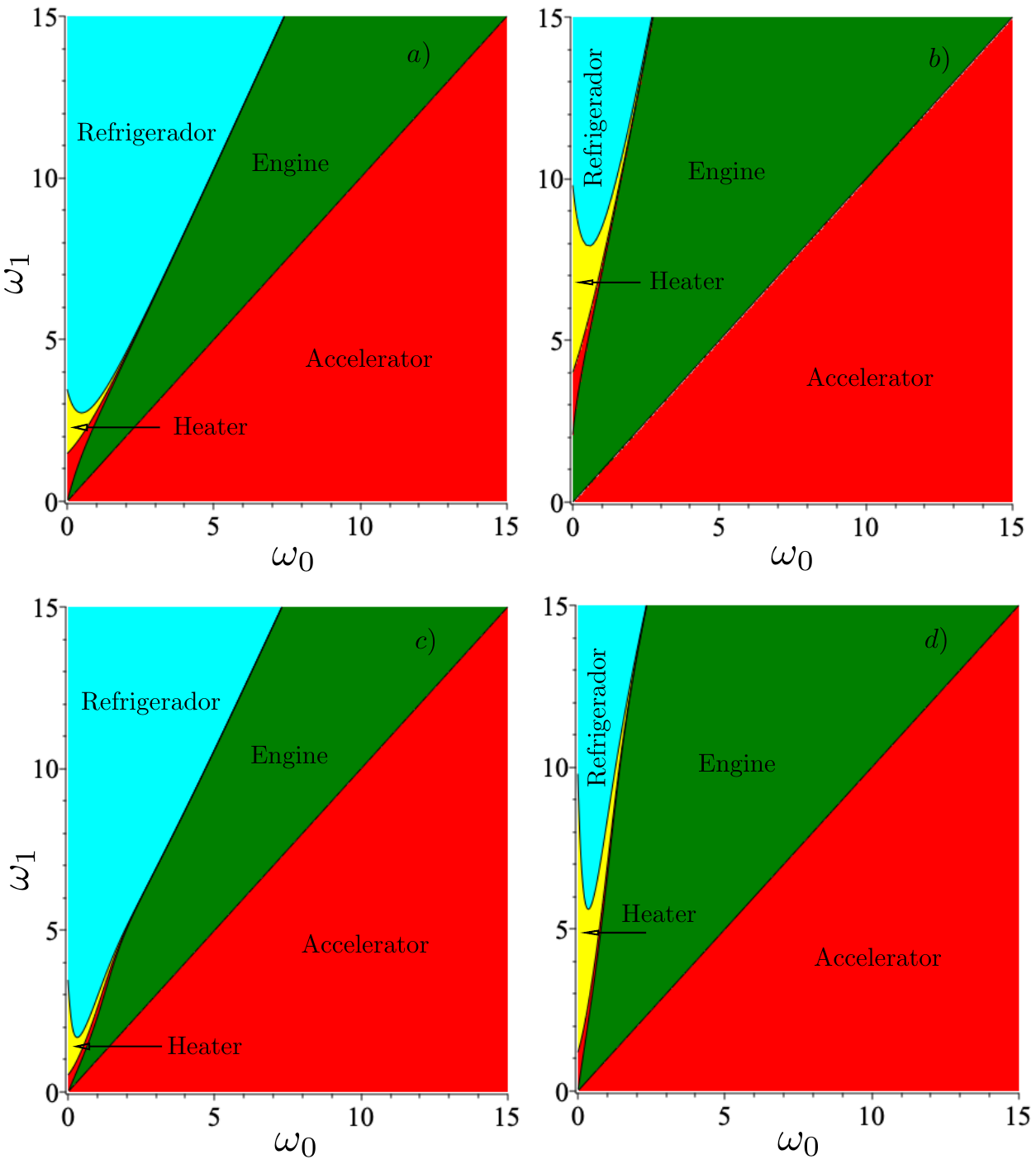} \caption{Operational modes of the quantum Otto cycle in the $\omega_{0}-\omega_{1}$ plane for different parameter configurations. a) $r=1$,
$T_{c}=1$, and $T_{h}=2$. b) $r=1$, $T_{c}=1$, and $T_{h}=5$.
c) $r=3$, $T_{c}=1$, and $T_{h}=2$. d) $r=3$, $T_{c}=1$,
and $T_{h}=5$. The coupling parameter was fixed at $g=1$.}
\label{fig:otto}
\end{figure}

Figure \textcolor{violet}{\ref{fig:otto}} shows the operational-mode
diagrams of the quantum Otto cycle in the $(\omega_{0},\omega_{1})$
plane. The colors distinguish heat engine (green), refrigerator (cyan),
thermal accelerator (red), and heater (yellow) behavior. In all cases,
the frequency applied to the right-qubit is modulated between $\omega_{0}$
and $\omega_{1}$, while the frequency on the left-qubit is fixed.

In panel \textcolor{violet}{\ref{fig:otto}}(a), with $T_{c}=1$,
$T_{h}=2$, and $r=1$, the heat-engine mode dominates the sector
$\omega_{1}>\omega_{0}$, where the expansion stroke enhances the
level spacing and allows positive work extraction. A refrigerator
region also appears in the same quadrant, but only for sufficiently
large frequencies $\omega_{1}\gtrsim2.8$, indicating that cooling
requires a minimum spectral response to thermalization. Thin heater
and accelerator bands reflect parameter ranges where spectral reshaping
and reservoir imbalance do not align favorably with work extraction.

The effect of increasing the thermal gradient is shown in panel \textcolor{violet}{\ref{fig:otto}}(b),
for $T_{h}=5$. The engine region expands significantly and extends
to larger frequency values, demonstrating that stronger thermal biases
enhance work output. In contrast, refrigeration becomes restricted
to very large values of $\omega_{1}$, meaning that cooling operation
is increasingly suppressed. Heater regions become more visible, signaling
stronger irreversibility under large temperature imbalance.

Panels \textcolor{violet}{\ref{fig:otto}}(c) and (d) illustrate the impact
of spectral asymmetry for $r=3$. For $T_{h}=2$ {[}panel
c{]}, the refrigeration region expands, confirming that stronger compression enhances
 population redistribution and, consequently, heat extraction from the
cold reservoir. The accelerator region is reduced, indicating that asymmetry
suppresses inefficient work-assisted conduction. When both effects
are combined, $T_{h}=5$ and $r=3$ in Fig. \textcolor{violet}{\ref{fig:otto}}(d),
the heat engine remains the dominant operational mode in $\omega_{1}>\omega_{0}$,
while the refrigerator window becomes narrow and heater behavior becomes
more prominent near the boundaries between regimes.

Overall, the Otto cycle exhibits a rich thermodynamic structure, with
its operational landscape controlled by the interplay between thermal
gradient and frequency asymmetry. Larger $T_{h}/T_{c}$ favors engine
performance, whereas stronger compression enhances refrigeration.
Accelerator and heater modes persist only in regions where the spectral
response to the stroke is insufficient to support efficient thermodynamic
conversion.

\subsubsection{Stirling Cycle}

\begin{figure}
\includegraphics[scale=0.45]{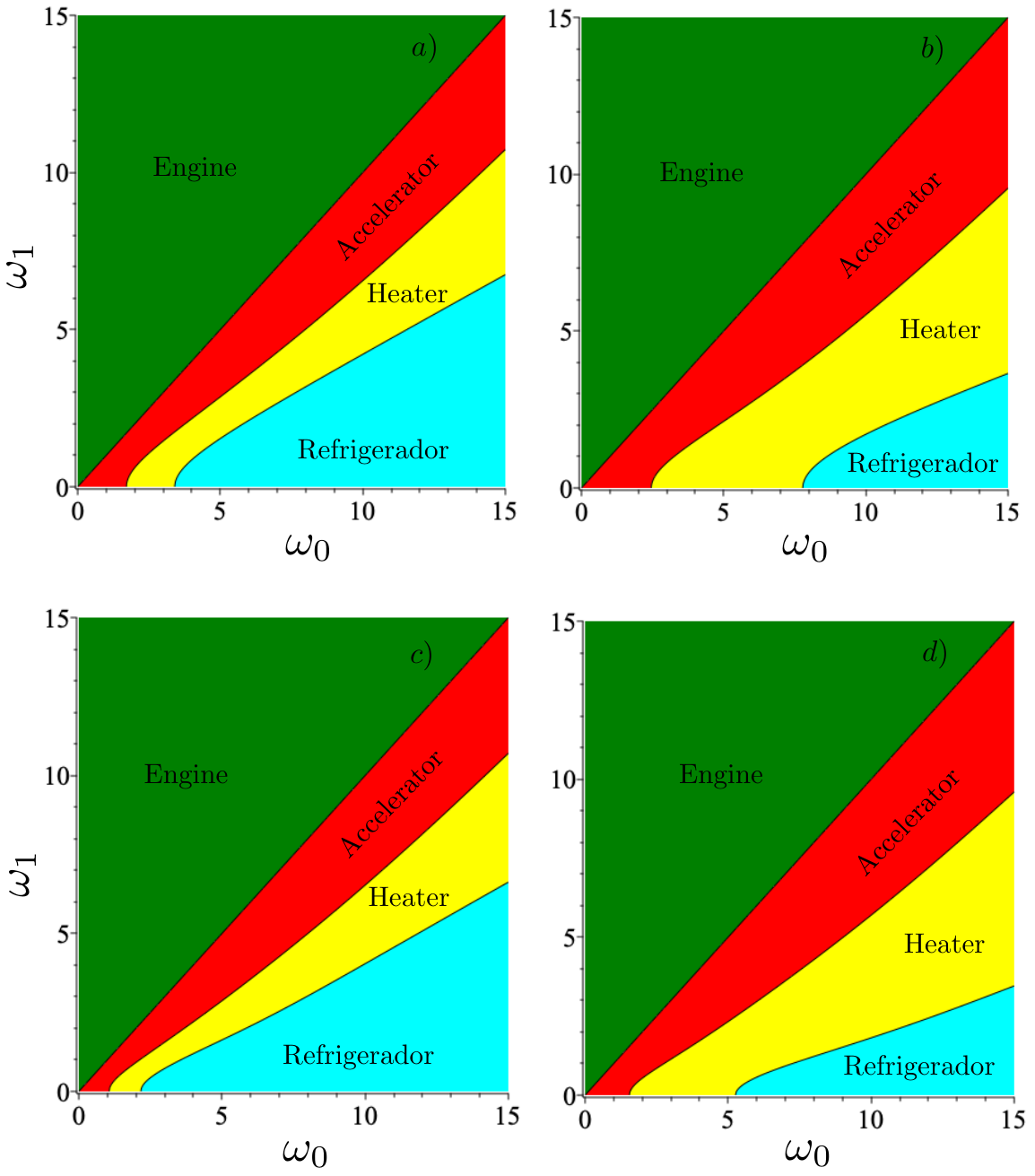} \caption{Operational modes of the quantum Stirling
cycle in the $\omega_{0}-\omega_{1}$ plane for different
parameter configurations. a)  $r=1$, $T_{c}=1$, and $T_{h}=2$.
b)  $r=1$, $T_{c}=1$, and $T_{h}=3$. c)  $r=2$, $T_{c}=1$,
and $T_{h}=2$. d)  $r=2$, $T_{c}=1$, and $T_{h}=3$. The coupling parameter
 was fixed at $g=1$.}
\label{fig:str-ram-1}
\end{figure}

\begin{figure}[t]
\includegraphics[scale=0.43]{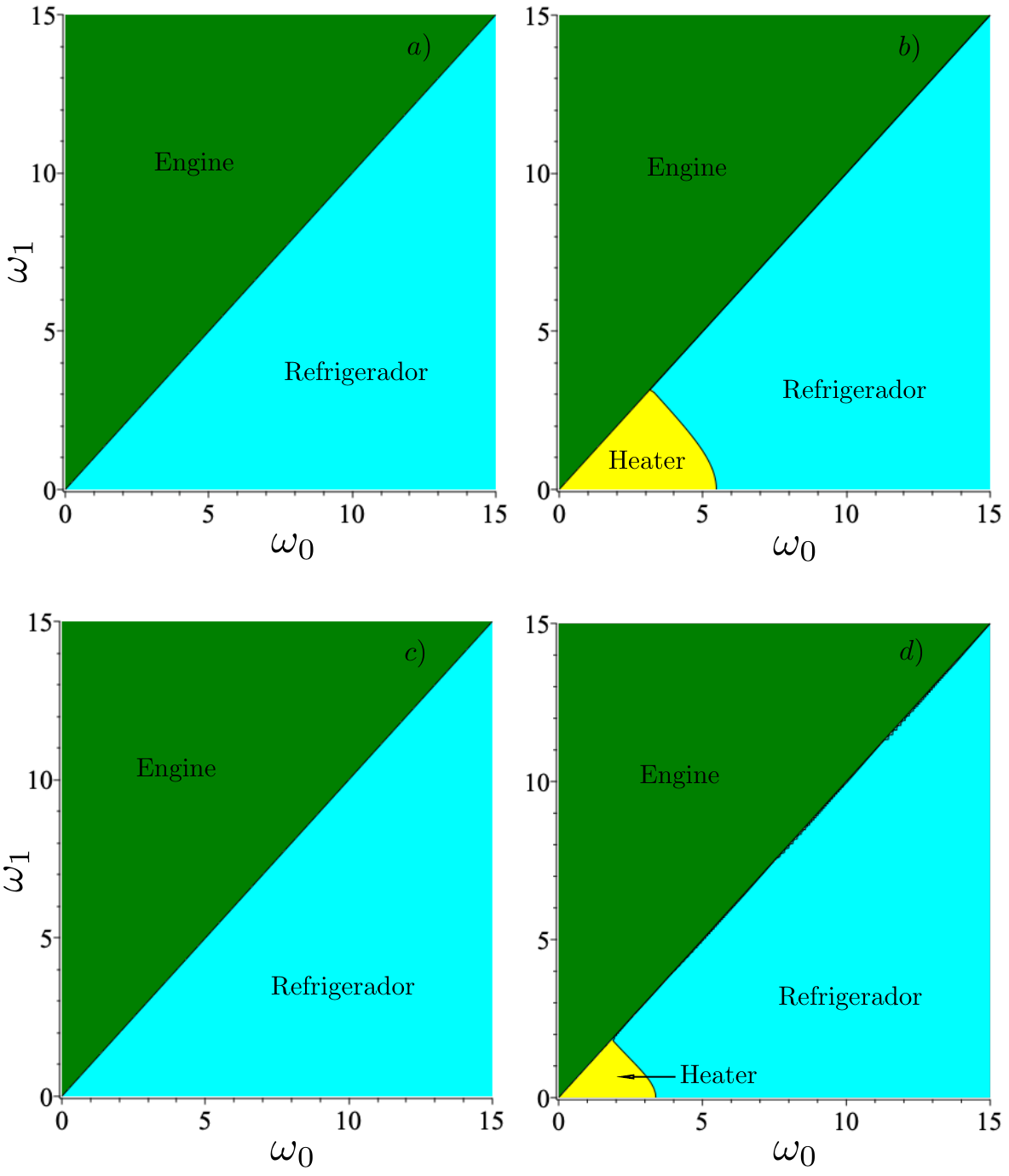} \caption{Operational modes of the quantum Stirling
cycle with a regenerator in the $\omega_{0}-\omega_{1}$ plane for different
parameter configurations. a) $r=1$, $T_{c}=1$, and $T_{h}=2$.
b) $r=1$, $T_{c}=1$, and $T_{h}=3$. c)  $r=2$, $T_{c}=1$,
and $T_{h}=2$. d)  $r=2$, $T_{c}=1$, and $T_{h}=3$.  The coupling parameter
 was fixed at $g=1$.}
\label{fig:str-reg}
\end{figure}

The Stirling protocol combines two isothermal and two constant-frequency
strokes, and its operational modes depend crucially on whether a regenerator
is present. Figure \textcolor{violet}{\ref{fig:str-ram-1}} depicts
the mode diagram without regeneration, while Fig.\textcolor{violet}{{}
\ref{fig:str-reg}} shows the corresponding behavior when an ideal
regenerator is included. In both cases the coupling is fixed at $g=1$,
and different panels explore the effect of temperature gradient and
frequency ratio.

\paragraph{Operation without regenerator}

Figure \textcolor{violet}{\ref{fig:str-ram-1}} presents the operational-mode
diagrams of the Stirling cycle in the $(\omega_{0},\omega_{1})$ plane.
Four operating regimes are observed: heat engine (green), refrigerator
(cyan), thermal accelerator (red), and heater (yellow).

In panels \textcolor{violet}{\ref{fig:str-ram-1}}(a) and (b), the parameters
are fixed at $g=1$, $r=1$, and $T_{c}=1$. For moderate thermal
bias, $T_{h}=2$ {[}panel \textcolor{violet}{\ref{fig:str-ram-1}}(a){]},
cooling operation requires sufficiently large frequencies with $\omega_{0}>\omega_{1}$
and $\omega_{0}\gtrsim3.5$. Increasing the thermal gradient to $T_{h}=3$
{[}panel \textcolor{violet}{\ref{fig:str-ram-1}}(b){]} shifts the onset
of refrigeration to significantly larger values of $\omega_{0}$,
while heater behavior expands near the boundaries. In both cases,
engine operation dominates for $\omega_{1}>\omega_{0}$, where the
stroke increases the hybridized energy gap.

Panels \textcolor{violet}{\ref{fig:str-ram-1}}(c) and (d) illustrate
the effect of spectral asymmetry for $r=2$. For $T_{h}=2$ {[}panel \textcolor{violet}{\ref{fig:str-ram-1}}(c){]}, refrigeration becomes accessible at lower frequencies, $\omega_{0}\gtrsim2.1$,
showing that stronger compression improves heat extraction from the
cold bath. When $T_{h}=3$ {[}panel \textcolor{violet}{\ref{fig:str-ram-1}}(d){]}, refrigeration again shifts
to larger $B_{0}$, accompanied by a broader heater region. Overall,
without regeneration the Stirling cycle displays a strong dependence
on the thermal gradient, and refrigeration is only achievable when
the spectral response to the isothermal strokes is sufficiently large.

\paragraph{Operation with regenerator}

Figure\textcolor{violet}{{} \ref{fig:str-reg}} displays the operational
maps of the Stirling cycle when a regenerator is included. The presence
of the regenerator eliminates the constant-frequency heat exchange
with the reservoirs, ensuring that heat transfer occurs only during
the isothermal strokes. In panel\textcolor{violet}{{} \ref{fig:str-reg}}(a),
for $r=1$, $T_{c}=1$, and $T_{h}=2$, only heat-engine and refrigerator
modes are present. The boundary between them coincides with the line
$\omega_{0}=\omega_{1}$. The absence of heater and accelerator regions
indicates that regeneration avoids dissipation into both reservoirs.
Increasing the thermal gradient to $T_{h}=3$ {[}panel\textcolor{violet}{{}
\ref{fig:str-reg}}(b){]} produces a small heater region for $\omega_{0}>\omega_{1}$
and small frequencies, reflecting the onset of irreversibility when
the input heat exceeds what can be stored and reused internally. When
effect of spectral asymmetry is increased to $r=2$ {[}panel \textcolor{violet}{{}
\ref{fig:str-reg}}(c){]}, the heater region is suppressed again and
only engine and refrigerator modes remain accessible, showing that
compression compensates for the thermal imbalance.

Finally, in panel\textcolor{violet}{{} \ref{fig:str-reg}}(d), combining
$r=2$ and $T_{h}=3$ reintroduces a narrow heater strip near small
$\omega_{0}$. Even so, the operational boundaries between engine
and refrigerator modes remain nearly unchanged. This demonstrates
that regeneration stabilizes Stirling operation and strongly reduces
dissipative behavior, even under large frequency asymmetry and pronounced
temperature bias.

\subsection{Efficiency and COP maps}

While the operational-mode diagrams identify where a cycle functions
as engine or refrigerator, the performance inside each region depends
on how effectively frequency modulation reshapes the energy spectrum
relative to the thermal gradient. Figures \textcolor{violet}{\ref{fig:carnot-eff}}-\textcolor{violet}{{}
\ref{fig:str-reg-ef}} display the efficiency for the Carnot and Otto
cycles, and the normalized coefficient of performance $\kappa$ for
the Stirling protocols, across the $(\omega_{0},\omega_{1})$ parameter
plane. These maps reveal how the Raman-induced spectral sensitivity
governs the conversion of thermal resources into useful work or cooling
power.

\subsubsection{Carnot efficiency}

\begin{figure}
\includegraphics[scale=0.4]{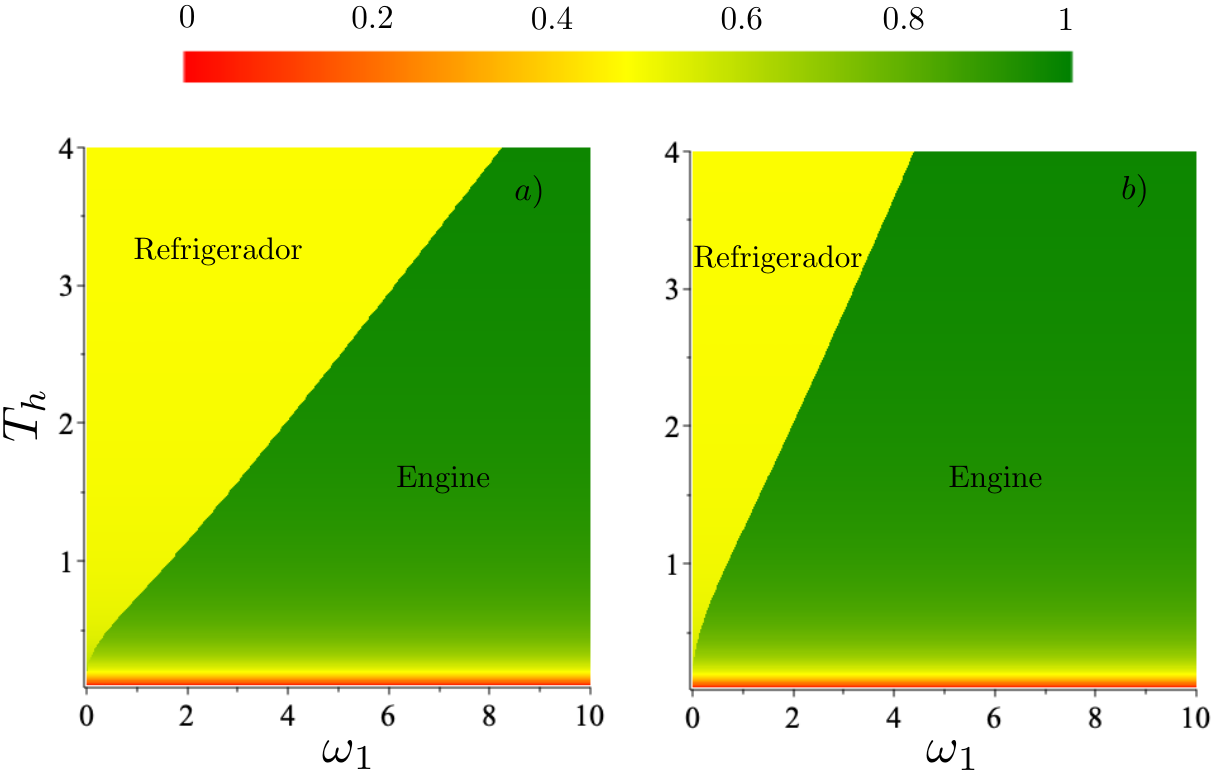}\caption{Normalized Carnot performance in the $\omega_{1}-T_{h}$
plane. a) $r=1$. b) $r=3$. The parameters are fixed at $T_{c}=0.1$,
$\omega_{0}=0$, and $g=1$.}
\label{fig:carnot-eff}
\end{figure}

Figure\textcolor{violet}{{} \ref{fig:carnot-eff}} shows the normalized
Carnot performance in the $(\omega_{1},T_{h})$ plane for two values
of frequency asymmetry. The normalization ensures that $\kappa=1$
corresponds to the ideal Carnot efficiency $1-T_{c}/T_{h}$, while
$\kappa=0$ indicates vanishing efficiency or COP. For $r=1$ {[}panel\textcolor{violet}{{}
\ref{fig:carnot-eff}}(a){]}, the operation is divided into heat engine
and refrigerator regions. The engine mode displays consistently high
efficiency, with $\kappa\approx0.9$ across a broad range of $T_{h}$,
indicating that frequency modulation is well aligned with entropy
reshaping during the isothermals. The refrigerator domain exhibits
more moderate performance, $\kappa\approx0.5$, and is confined to
smaller values of $\omega_{1}$ and $T_{h}$. When the spectral asymmetry
increases to $r=3$ {[}panel\textcolor{violet}{{} \ref{fig:carnot-eff}}(b){]},
the qualitative structure remains the same, but the engine region
expands while the refrigerator domain becomes narrower. This indicates
that a stronger compression ratio facilitates work extraction even
for weaker frequency strokes. In both cases, the contour lines remain
almost horizontal, showing that the Carnot performance is controlled
predominantly by the thermal gradient and is only weakly sensitive
to the absolute frequency value within the parameter range considered.

Overall, the Carnot protocol demonstrates that whenever engine operation
is thermodynamically allowed, it proceeds with efficiency close to
the reversible bound. The main limitation stems from progressive shrinking
of the admissible operation region as the frequency ratio or the thermal asymmetry increases.

\subsubsection{Otto efficiency}

\begin{figure}
\includegraphics[scale=0.4]{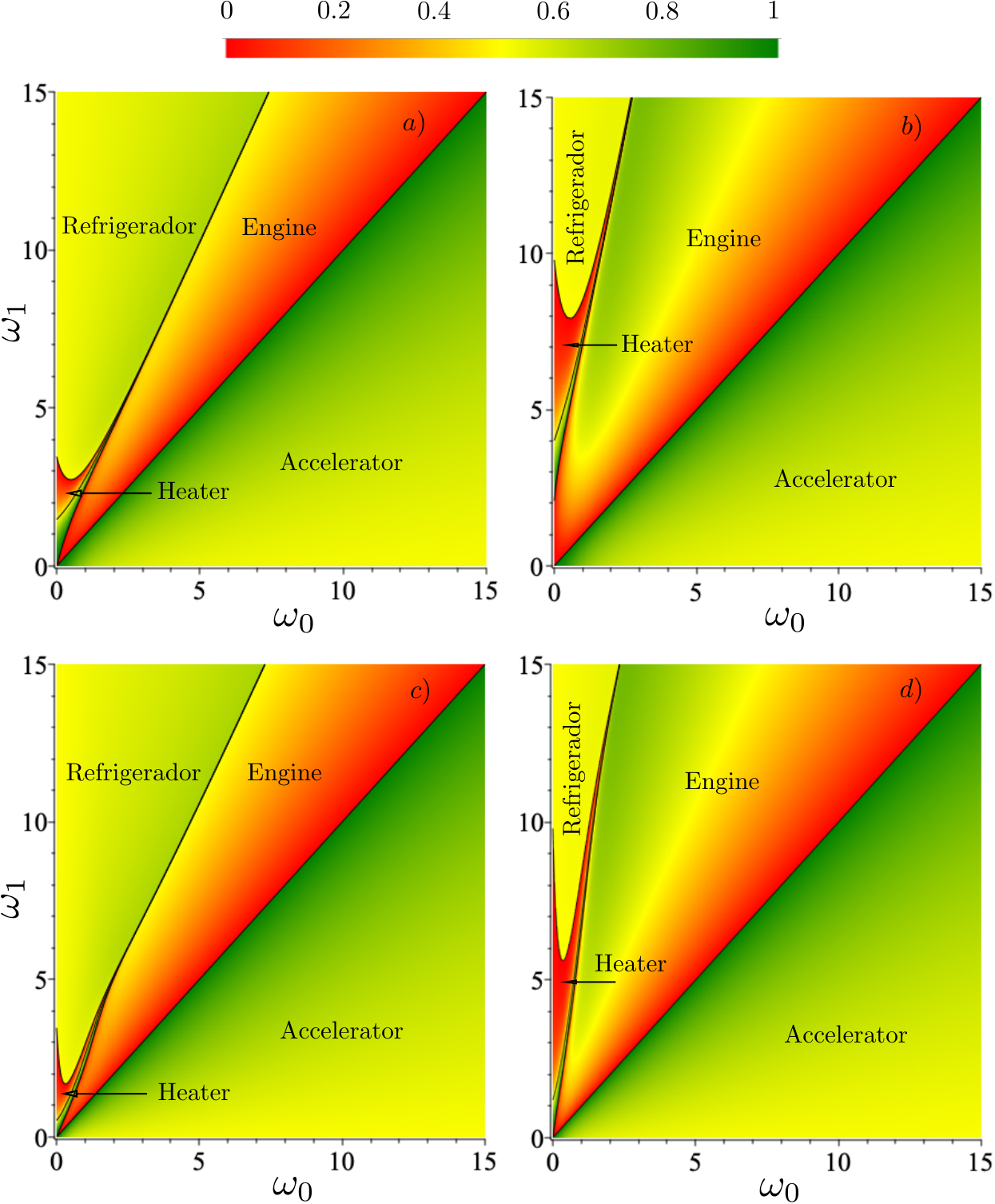} \caption{Thermal efficiency of the quantum Otto cycle in the $\omega_{0}$-$\omega_{1}$
plane. a) $r=1$,
$T_{c}=1$, and $T_{h}=2$. b) $r=1$, $T_{c}=1$, and $T_{h}=5$.
c) $r=3$, $T_{c}=1$, and $T_{h}=2$. d) $r=3$, $T_{c}=1$,
and $T_{h}=5$. The coupling parameter was fixed at $g=1$.}
\label{fig:otto-eff} 
\end{figure}

Figure\textcolor{violet}{{} \ref{fig:otto-eff}} shows the normalized
performance $\kappa$ of the Otto cycle in the $(\omega_{0},\omega_{1})$
plane for different thermal gradients and frequency asymmetry. For
heat‐engine operation, $\kappa$ coincides with the thermodynamic
efficiency $\eta$. In the refrigerator, heater, and accelerator modes,
$\kappa$ corresponds to the normalized COP, enabling direct comparison
across all regimes.

For $r=1$ and $T_{h}=2$ {[}panel\textcolor{violet}{{} \ref{fig:otto-eff}}(a){]},
the heat‐engine regime dominates the sector $\omega_{1}>\omega_{0}$,
where the expansion stroke increases the level splitting. Performance
is modest, $\eta\lesssim0.5$, but increases near the boundary with
the refrigerator region, reflecting enhanced thermodynamic response
at the onset of cooling operation. Refrigeration requires sufficiently
strong hybridization and appears only at larger frequencies. Accelerator
and heater modes remain confined to small pockets where the direction
of spectral reshaping does not align with the thermal bias.

A stronger thermal gradient, $T_{h}=5$ {[}panel\textcolor{violet}{{}
\ref{fig:otto-eff}}(b){]}, substantially expands the engine region,
boosting efficiency up to $\eta\approx0.8$, while refrigeration becomes
restricted to very large $\omega_{1}$. This indicates that strong
temperature bias favors work extraction but makes cooling more difficult.
Heater behavior increases near the operational boundaries due to growing
irreversibility.

Panels\textcolor{violet}{{} \ref{fig:otto-eff}}(c) and (d) show the
effect of spectral asymmetry, $r=3$. At moderate thermal bias {[}$T_{h}=2$,
panel\textcolor{violet}{{} \ref{fig:otto-eff}}(c){]}, refrigeration
expands toward lower frequencies, demonstrating that compression enhances
population reshaping and enables more efficient cooling. Heater
operation is significantly suppressed. When both parameters are large,
$r=3$ and $T_{h}=5$ {[}panel\textcolor{violet}{{} \ref{fig:otto-eff}}(d){]},
the heat‐engine region remains dominant with high performance, while
refrigeration is confined to a narrow strip at large frequencies, while
heater regions become more pronounced near the low-frequency boundaries, where the coefficient $\kappa$ attains small values.

In summary, Otto performance is governed by the interplay between
spectral reshaping and thermal bias: increasing $T_{h}/T_{c}$ favors
work extraction, whereas increasing $r$ favors refrigeration. Accelerator
and heater behavior appear when neither condition is sufficiently
met for efficient operation.

\subsubsection{Stirling cycles with and without regeneration}

\begin{figure}
\includegraphics[scale=0.43]{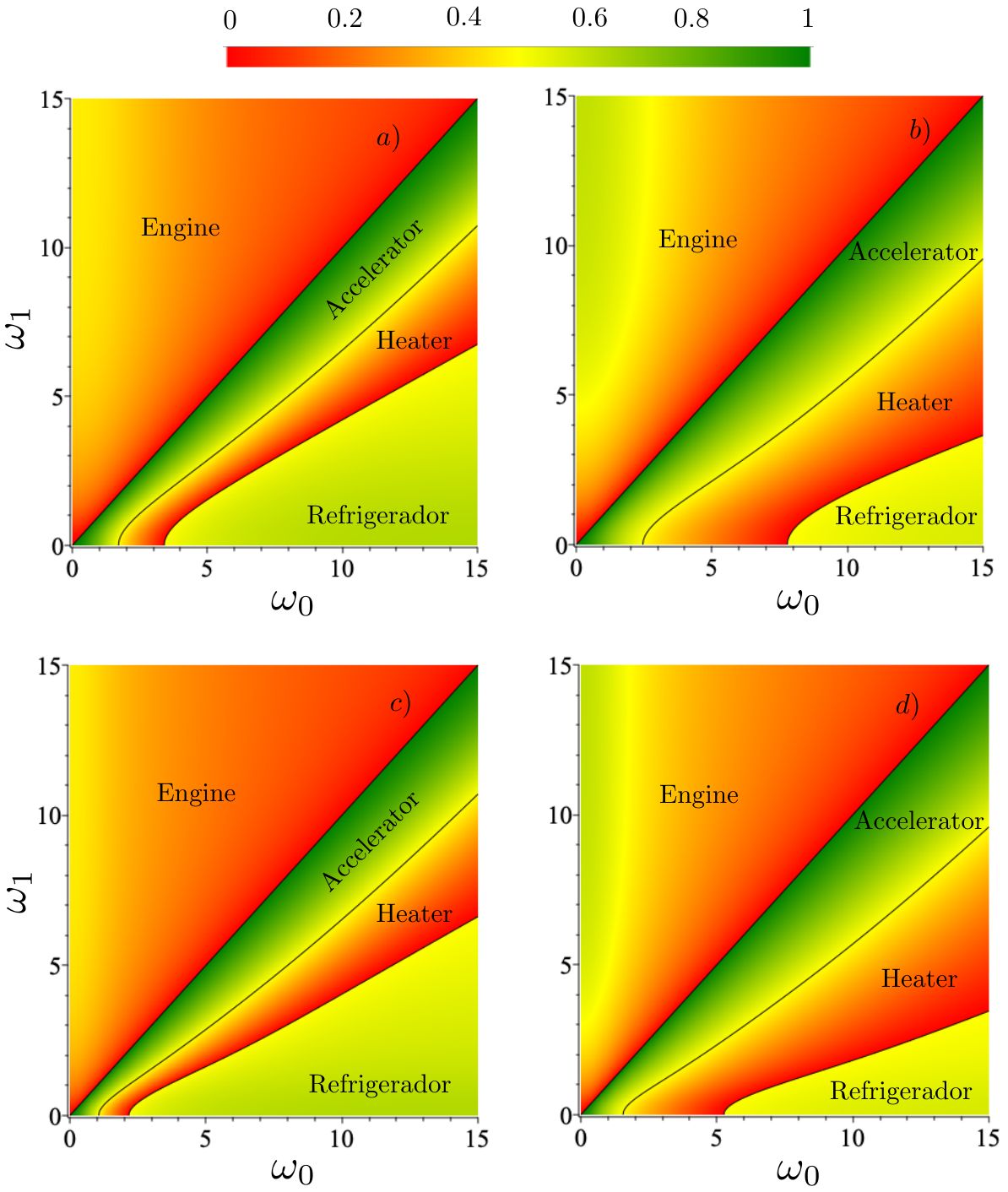} \caption{Thermal efficiency of the quantum Stirling cycle in the $\omega_{0}$-$\omega_{1}$
plane. a) $r=1$, $T_{c}=1$, and $T_{h}=2$. b) $r=1$, $T_{c}=1$, and $T_{h}=3$. c) $r=2$, $T_{c}=1$, and
$T_{h}=2$. d) $r=2$, $T_{c}=1$, and $T_{h}=3$. The coupling parameter was fixed at $g=1$.}
\label{fig:ef-sti-sre}
\end{figure}

\begin{figure}
\includegraphics[scale=0.42]{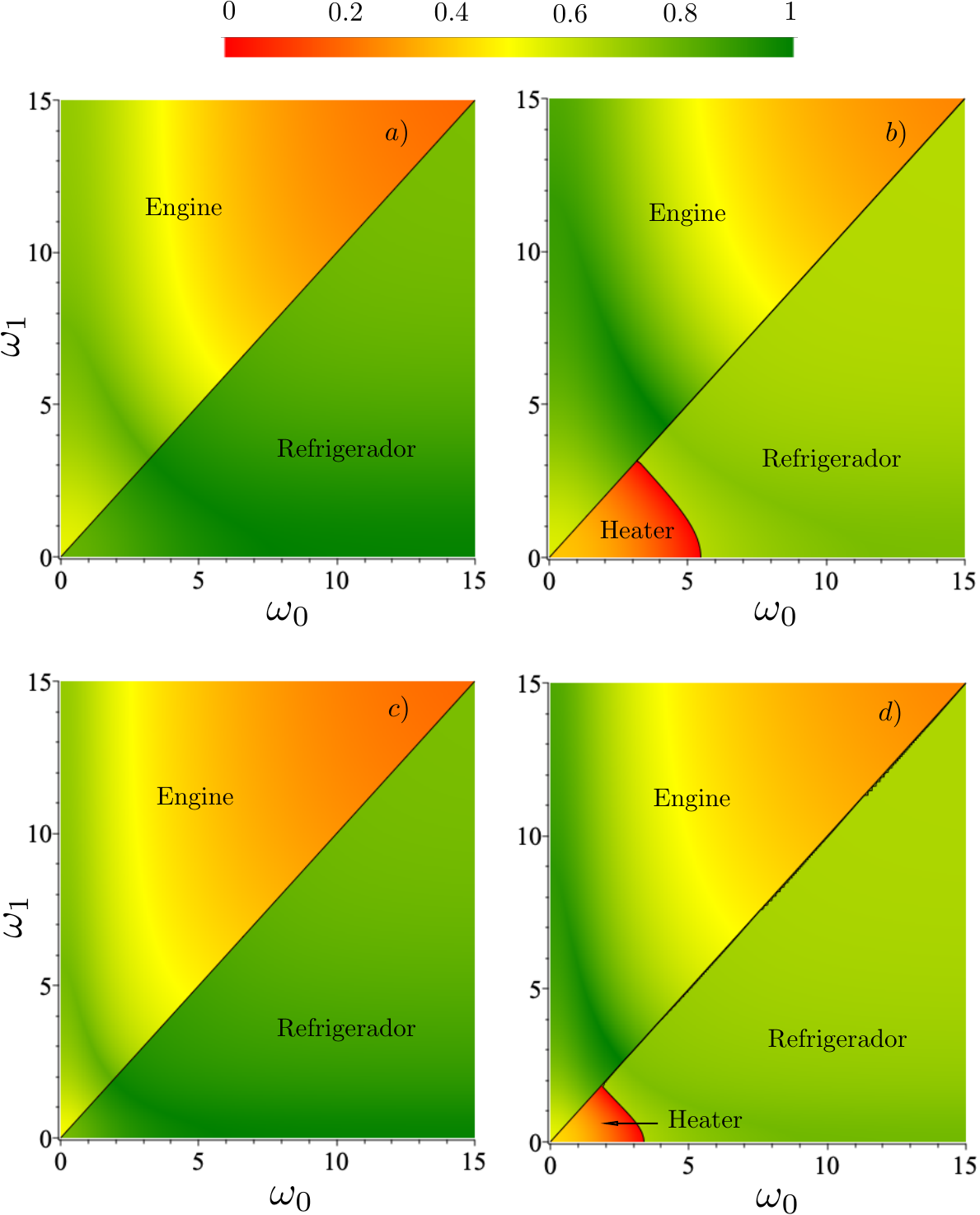} \caption{Thermal efficiency of the quantum Stirling cycle in the $\omega_{0}$-$\omega_{1}$
plane. a) $r=1$, $T_{c}=1$, and $T_{h}=2$. b) $r=1$ $T_{c}=1$,
and $T_{h}=3$. c) $r=2$, $T_{c}=1$, and $T_{h}=2$. d) $r=2$,
$T_{c}=1$, and $T_{h}=3$. The parameter $g$ was fixed at $g=1$.}
\label{fig:str-reg-ef}
\end{figure}

Figures \textcolor{violet}{\ref{fig:ef-sti-sre}} and \textcolor{violet}{\ref{fig:str-reg-ef}}
present the performance maps of the Stirling cycle, without and with
an ideal regeneration process included. These diagrams reveal that
Stirling behavior is governed primarily by how efficiently thermal
population reshaping is exploited during the two isothermal strokes
and how much heat is lost during the constant-frequency branches.

\paragraph{Stirling without regeneration}

Figure\textcolor{violet}{{} \ref{fig:ef-sti-sre}} shows the normalized
performance $\kappa$ of the Stirling cycle without regeneration in
the $(\omega_{0},\omega_{1})$ plane. As in the Otto case, $\kappa$
equals the engine efficiency in the $\omega_{1}>\omega_{0}$ region
and represents the normalized COP in the refrigerator and accelerator
regimes. For $g=1$, $r=1$, and $T_{h}=2$ {[}panel\textcolor{violet}{{}
\ref{fig:ef-sti-sre}}(a){]}, engine operation dominates for $\omega_{1}>\omega_{0}$
with moderate performance $\eta\approx0.5$. Refrigeration is only
feasible for larger $\omega_{0}$, where hybridization is strong enough
to support heat extraction. Accelerator behavior exhibits high $\kappa$
values due to large entropy reshaping under quasi-degenerate conditions,
whereas the heater mode remains inefficient. Raising the thermal gradient
to $T_{h}=3$ {[}panel\textcolor{violet}{{} \ref{fig:ef-sti-sre}}(b){]}
enhances the engine efficiency and shifts refrigeration to higher
$\omega_{0}$. Heater regions expand because dissipation increases
when the internal heat released during the constant-frequency strokes
is dumped directly into the baths. Increasing the frequency asymmetry
to $r=2$ {[}panels\textcolor{violet}{{} \ref{fig:ef-sti-sre}}(c) and
(d){]} favors refrigeration by intensifying the spectral response of
the strokes, enabling cooling at lower frequencies. The engine regime maintains
similar performance, while the heater mode remains restricted to regions of low efficiency. Under conditions of high $r$ and high $T_{h}$ {[}panel \textcolor{violet}{{} \ref{fig:ef-sti-sre}}(d){]},
the engine efficiency is significantly improved; however, refrigeration becomes
limited, demonstrating a more pronounced competition between frequency ratio 
and thermal gradient.

Overall, without regeneration the performance is strongly limited
by the heat dumped into the reservoirs during constant-frequency strokes,
enhancing heater behavior and reducing cooling efficiency.

\paragraph{With regeneration}

Figure \ref{fig:str-reg-ef} displays the performance of the Stirling
cycle including an ideal regenerator. In this case, heat exchanged
during the constant-frequency strokes is internally stored and reused,
eliminating an important source of dissipation. For $r=1$ and $T_{h}=2$
{[}panel \ref{fig:str-reg-ef}(a){]}, the heater mode disappears entirely.
High performance is achieved in both engine $(\omega_{1}>\omega_{0})$
and refrigerator $(\omega_{1}<\omega_{0})$ regions, demonstrating
that regeneration preserves useful heat fluxes and greatly improves
reversibility. When $T_{h}$ is increased to 3 {[}panel \ref{fig:str-reg-ef}(b){]},
engine efficiency rises to $\eta\approx0.8$, while a small frequency
heater strip appears due to growing thermal imbalance. Cooling performance
slightly decreases but still remains significantly better than without
regeneration. Frequency asymmetry enhances these advantages. For $r=2$
with $T_{h}=2$ {[}panel \ref{fig:str-reg-ef}(c){]}, engine efficiencies
as high as $\eta\approx0.9$ are achieved even for relatively small
frequencies, and refrigeration is highly efficient for $\omega_{1}<\omega_{0}$.
Under both increased $r$ and strong thermal gradient {[}panel \ref{fig:str-reg-ef}(d){]},
the engine remains the dominant mode with optimal performance, while
only a very narrow heater region persists.

In summary, regeneration suppresses dissipation, expands efficient
modes, and enables high performance even under asymmetry and strong
drive. It is thus the key mechanism that unlocks the full thermodynamic
potential of the Raman-coupled working medium.

\subsection{Distinct thermodynamic signatures}

The operational-mode diagrams and performance maps reveal robust thermodynamic
features that persist across Carnot, Otto, and Stirling cycles. These
features are not tied to a specific protocol, but originate from how
frequency tuning of the right qubit reshapes the spectrum and controls
population redistribution in the Raman-coupled working medium. As
a result, similar qualitative patterns emerge across all cycles.

\paragraph{Internal heat recovery.}

The figures show that internal heat management critically determines
the accessible operation modes. In the absence of regeneration, regions
appear where energy exchange with the reservoirs does not translate
into useful work or cooling, leading to accelerator or heater behavior
near the boundaries between productive regimes. Introducing a regenerator
strongly suppresses these regions by recycling heat exchanged during
constant-frequency strokes. This demonstrates that the dominant inefficiencies
observed in the maps arise from external dissipation rather than from
the spectral structure itself.

\paragraph{Performance degradation in extreme regimes.}

The maps further indicate that extreme parameter regimes degrade thermodynamic
performance. Large thermal biases favor irreversible population reshaping,
while very small driving frequencies limit the ability of spectral
modulation to induce effective energy conversion. In both cases, extended
low-performance regions emerge, corresponding to heater-dominated
operation. These regimes therefore represent intrinsic limits of frequency-based
control, rather than artifacts of a particular cycle.

\paragraph{Frequency-driven enhancement.}

Finally, all figures exhibit extended regions of enhanced performance
associated with strong spectral responsiveness to frequency tuning.
In these regions, modest frequency variations induce significant population
redistribution, enhancing entropy changes along isotherms, work extraction
in Otto cycles, and refrigeration performance in Stirling protocols.
When regeneration is included, this response-driven mechanism dominates
the efficiency maps, as dissipative losses are minimized and the available
spectral sensitivity is fully exploited.

Overall, these results show that the thermodynamic behavior of the
Raman-coupled two-qubit machine is governed primarily by population
sensitivity to frequency control, rather than by the detailed structure
of the cycle. This explains the recurring operational patterns observed
across different thermodynamic protocols and highlights frequency
tuning as the central organizing principle underlying the reported
results.

\section{Conclusions}\label{sec:5}

We have analyzed the thermodynamic performance of a quantum heat machine
whose working medium consists of two qubits coupled through a Raman-type
exchange interaction. By examining Carnot, Otto, and Stirling operation
(with and without regeneration), we mapped in detail the operational
modes that emerge when the right-qubit frequency is driven between
two values $\omega_{0}$ and $\omega_{1}$. The resulting phase diagrams
reveal sharp boundaries separating heat engine, refrigerator, accelerator,
and heater behavior, determined by the interplay between frequency
asymmetry, Raman mixing, and thermal bias.

In the Carnot cycle, reversible operation is achieved only within
a narrow region, close to the optimal frequency ratios. Increasing the temperature
gradient broadens the refrigeration domain and restricts engine operation,
even though the maximum attainable efficiency rises. The Otto cycle
exhibits a richer multimode structure, with all four operational regimes
coexisting over large portions of parameter space. Enhanced spectral
asymmetry favors refrigeration, while strong thermal drive improves
engine performance but suppresses accelerator behavior.

For the Stirling cycle, the inclusion of a regenerator markedly improves
reversibility by internally recycling heat released in the constant-frequency
strokes. As a result, heater and accelerator domains are strongly
suppressed, and both engine and cooling modes reach near-ideal performance.
Under strong coupling and high thermal bias, efficiencies approach
$\kappa\approx0.9$, outperforming the other cycles while remaining
robust across a wide range of parameters.

Overall, the Raman-induced level hybridization introduces a controllable
left-right asymmetry that enables continuous transitions between distinct
thermodynamic behaviors using only frequency tuning. These results
demonstrate that Raman-coupled qubit architectures represent highly
versatile nanoscale working media, capable of optimized energy conversion
in diverse operational regimes and well suited for the design of next-generation
quantum thermal machines.
\begin{acknowledgments}
O. R. and M. Rojas thanks CNPq and FAPEMIG for partial financial support.
M. Rojas acknowledges CNPq Grant 311565/2025-5. 
\end{acknowledgments}

\end{document}